\documentclass[showpacs,aps,pre,twocolumn]{revtex4}
\usepackage{graphicx,graphics}
\begin {document}
\title{\bf{A New Damping Mechanism in Non-linear Bubble Dynamics}}
\author{Ahmad Moshaii$^{1,3}$}\email{moshaii@mehr.sharif.edu} \author{Rasool Sadighi-Bonabi $^{1,2}$}  \author{Mohammd Taeibi-Rahni$^{4}$}
\address{$^1$ Department of Physics, Sharif University of Technology, P.O. Box:11365-9161, Tehran, I.R. Iran}
\address{$^2$ Bonab Research Center, P.O. Box:56515-196, Bonab, Azarbayejan Province, I.R. Iran}
\address{$^3$ Institute for Studies in Theoretical Physics and Mathematics, P.O. Box:19395-5531, Tehran, I.R. Iran}
\address{$^4$ Department of Aerospace Engineering, Sharif University of Technology, P.O. Box:11365-9161, Tehran, I.R. Iran}
\begin{abstract}

Non-linear equations of radial motion of a gas bubble in a
compressible viscous liquid have been modified considering
effects of viscosity and compressibility more complete than all
previous works. A new set of equations has been derived including
new terms resulted from consideration of the viscosity and
compressibility not only at the bubble interface, but also in the
bulk of liquid. The new equations are two non-linear coupled
equations, which can not be merged into one equation unlike all
previously derived equations. Numerical calculations have been
performed considering effects of heat and mass transfer at the
bubble interface. The results indicate that the new terms exhibit
an important damping role at the collapse, so that their
consideration dramatically weakens the bubble rebounds after the
collapse. Dependence of this new damping mechanism to amplitude
and frequency of the deriving pressure has been investigated.
\end{abstract} \maketitle

\section{Introduction}
When a small isolated gas bubble, immersed in a liquid,
experiences a high amplitude spherical sound field, it grows and
contracts non-linearly. Description of the dynamics of such
non-linear motion is an old challenging problem. The complexities
of the problem arise from the effects of heat conduction, mass
diffusion, compressibility, viscosity, and energy losses involved
in damped oscillations of the bubble. So far, many people have
investigated this problem, each concentrating on different
aspects of its complexities. However, a rather complete
description has not been presented yet.

Lord Rayleigh \cite{Rayleigh:1917} was the first who analytically
solved the collapse of an empty bubble in an incompressible
liquid. Plesset \cite{Plesset:1949} subsequently studied the
dynamics of a vapor-filled bubble in a time-dependent pressure
field. Noltingk and Neppiras \cite{Noltingk:1951} were the first
people derived a differential equation for the gas bubble motion
under the influence of such pressure field. The extension of this
equation to the bubble motion in a compressible liquid has been
studied by many authors; Herring \cite{Herring:1941}, Trilling
\cite{Trilling:1952}, Gilmore \cite{Gilmore:1952}, Keller and
Kolodner \cite{Keller:1956}, Hickling and Plesset
\cite{Hickling:1964}, Jahsman \cite{Jahsman:1968}, Flynn
\cite{Flynn:1975}, Lastman and Wentzell \cite {Lastman:1979},
Keller and Miksis \cite{Keller:1980}. On the other hand, heat
conduction effects were presented in the works of Hickling
\cite{Hickling:1963}, Fujikawa and Akumatsu \cite{Fujikawa:1980},
and Yasui \cite{Yasui:1997}. In addition, the works of
L\"{o}fstedt \textit{et al.} \cite{fstedt:1993}, Nigmatulin
\textit{et al.} \cite{Nigmatulin:2000}, and Brujan
\cite{Brujan:2001} are of the subsequent papers addressing this
problem.

Between all previous works, the most complete description of the
bubble dynamics equations was presented by Prosperetti and Lezzi
\cite{Prosperetti:1987}. They used a singular-perturbation method
of the bubble-wall Mach number and derived the following
one-parameter family of equations describing the bubble motion in
the first order approximation of compressibility
\cite{Prosperetti:1987}:
\begin{eqnarray}
\label{eq1} \left({1-(\eta + 1)\frac{\dot{R}}{C}}
\right)\!R\ddot{R}+\frac{3}{2}\left({1-\frac{1}{3}(3\eta +
1)\frac{\dot {R}}{C}}\right)\!\dot{R}^2 = \nonumber \\
{\frac{R}{\rho C}\frac{d}{dt}\left( {P_l - P_\infty }
\right)\!+\!\left({1+(1-\eta)\frac{\dot {R}}{C}} \
\right)\!\!\!\left( \frac{P_l - P_\infty }{\rho}
\right)},\nonumber \\
\end{eqnarray}
\noindent where, $R$, $C$, $P_{\infty }$, and $\rho$ are the
bubble radius, liquid sound speed, liquid pressure at infinity,
and liquid density, respectively. Also, $\eta $ is an arbitrary
parameter. Equation (\ref{eq1}) must be supplemented by a
boundary condition equation at the bubble interface to relate the
liquid pressure, $P_{l}$, to the gas pressure inside the bubble.
Like all previous authors, Prosperetti and Lezzi
\cite{Prosperetti:1987} used the following incompressible
equation for this purpose:
\begin{equation}
\label{eq2} P_l = P_g - 4\mu \frac{\dot {R}}{R}-\frac{2\sigma
}{R},
\end{equation}
\noindent where, $P_{g}$, $\mu$, and $\sigma$ are the gas pressure
at the bubble interface, liquid viscosity, and surface tension,
respectively. Most of the previously derived equations belong to
this single parameter family of equations, corresponding to
different values of $\eta$. Moreover, $\eta = 0$ yields results
in closest agreement with the numerical simulation of full partial
differential equations \cite{Prosperetti:1987}.

Two specific approximations have been used in the derivation of
Eq'ns. (\ref{eq1}) and (\ref{eq2}). The first approximation is
the negligence of the viscosity effects for the liquid motion
around the bubble, which has been used in the derivation of Eq'n.
(\ref{eq1}). In fact, Eq'n. (\ref{eq1}) has been derived from the
Euler equation, in which the viscosity is eliminated. Note that
the viscous term of Eq'n. (\ref{eq2}) has been resulted from the
liquid viscosity at the bubble interface, but not from the bulk
of liquid.

The second approximation is the incompressibility assumption of
the liquid and the gas at the bubble interface, which has been
used in the derivation of Eq'n. (\ref{eq2}). All of the effects
of the liquid compressibility in the work of Prosperetti and
Lezzi, as well as in all other previous works, have been resulted
from the liquid motion around the bubble, but not from the bubble
boundary condition equation. In fact, all previous authors, on
one hand took into account the compressibility of the liquid
motion around the bubble, but on the other hand neglected its
consideration at the bubble interface.

Although, the two mentioned approximations have been used in the
derivations of the existing bubble dynamics equations
\cite{Rayleigh:1917,Plesset:1949,Noltingk:1951,Herring:1941,Trilling:1952,
Gilmore:1952,Keller:1956,Flynn:1975,Jahsman:1968,Lastman:1979,fstedt:1993,
Nigmatulin:2000,Brujan:2001,Hickling:1963,Fujikawa:1980,Yasui:1997,Hickling:1964,Prosperetti:1987,Keller:1980},
but the applicability of these approximations for all times of the
bubble motion needs to be clarified.  Especially, at the end of
the collapse, when the bubble motion is significantly
compressible, these approximations may be inapplicable.

In this paper, by eliminating the above mentioned approximations,
we have originally modified the bubble dynamics equations. A new
set of equations have been derived including all effects of the
viscosity and compressibility of both the liquid and the gas.
These equations contain new terms resulted from the effects of
two coefficients of viscosity of both the liquid and the gas. The
influence of the added new terms arisen from the liquid viscosity
has been numerically investigated. The results clearly indicate
that the addition of the new terms considerably affects the
bubble motion at the collapse time and during the bubble rebounds.

\section{Derivation of the Bubble Dynamics Equations}

To derive the bubble dynamics equations, we assume that the motion
of the bubble interface and the infinitely extended surrounding
liquid are always spherically symmetric. Under this circumstance,
the continuity and the momentum equations are as follows
\cite{Aris:}:

\begin{equation}
\label{eq3}
\frac{1}{\rho }\left[ {\frac{\partial \;\rho }{\partial t} + u\frac{\partial
\,\rho }{\partial r}} \right] = - \frac{\partial \,u}{\partial r} -
\frac{2u}{r} = - \Delta ,
\end{equation}

\begin{eqnarray}
\label{eq4}
 \rho \left[ {\frac{\partial u}{\partial t} + u\frac{\partial u}{\partial
r}} \right] &=& - \frac{\partial p}{\partial r} + (\lambda + \mu
)\left[ {\frac{\partial }{\partial r}\left(
{\frac{1}{r^2}\frac{\partial }{\partial
r}\left( {r^2u} \right)} \right)} \right] \nonumber \\
 &+& \mu \left[ {\frac{1}{r^2}\frac{\partial }{\partial r}\left(
{r^2\frac{\partial u}{\partial r}} \right) - \frac{2u}{r^2}}
\right],
\end{eqnarray}
\noindent where, $\rho$, $u$, $p$, and $\Delta $ are density,
velocity, pressure, and divergence of the velocity, respectively.
Also, $\mu$ and $\lambda$ are first and second coefficients of
viscosity. The two brackets of the right hand side of Eq'n.
(\ref{eq4}) have the same differential forms. Hence:
\begin{equation}
\label{eq5} \rho\left( \frac{\partial u}{\partial t} +u
\frac{\partial u}{\partial r}\right)=-\frac{\partial p}{\partial
r}+(\lambda +2\mu)\frac{\partial \Delta}{\partial r}.
\end{equation}

\noindent Dividing Eq'n. (\ref{eq5}) by $\rho$ and integrating it
with respect to $r$ from the bubble interface to infinity at a
fixed instant of time and assuming that the two coefficients of
viscosity are constant, we get:
\begin{eqnarray}
\label{eq6} \int_{R}^{\infty}\!\frac{\partial u}{\partial t}
\;dr-\frac{\dot{R}^2}{2}=-\!\int_{R}^{\infty}\frac{1}{\rho}\;dp \;
+ (\lambda + 2\mu )\!\int_{R}^{\infty}\frac{1}{\rho}\;d\Delta,
\nonumber \\
\end{eqnarray}

\noindent where, the liquid velocity at the far field is assumed
to be sufficiently small. When the liquid density does not
strongly change, Eq'n. (\ref{eq6}) can be written as:
\begin{eqnarray}
\label{eq11} \int_{R}^{\infty}\frac{\partial u}{\partial t}dr
=\frac{\dot {R}^{2}}{2}-\frac{P_{\infty}-P_{l}}{\rho}
+\frac{\lambda+2\mu}{\rho}(\triangle_{\infty}-\triangle_{l}).
\nonumber \\
\end{eqnarray}

\noindent where, $\Delta_l$ and $\Delta_{\infty}$ are the
divergence of the liquid velocity at the bubble interface and at
infinity, respectively. The liquid pressure at the far field,
$P_{\infty}$, is:
\begin{equation}
\label{eq10} P_{\infty}(t)=P_0+P_a(t),
\end{equation}
\noindent where, $P_{0}$ and $P_{a}(t)$ are ambient and driving
pressures, respectively. From Eq'n. (\ref{eq3}) the divergence of
the velocity can be written as:
\begin{equation}
\label{eq12}
\triangle=-\frac{1}{\rho}\frac{d\rho}{dt}=-\frac{1}{\rho
c^{2}}\frac{dp}{dt},
\end{equation}

\noindent where the sound speed, $c$, which is assumed to be
constant in the liquid, is defined as: $c^2=dp/d\rho$. Thus, Eq'n.
(\ref{eq11}) becomes:
\begin{eqnarray}
\label{eq13} \int_{R}^{\infty}\frac{\partial u}{\partial t}dr
=\frac{\dot{R}^{2}}{2}-\frac{P_{\infty}-P_{l}}{\rho}
+\frac{\lambda+2\mu}{\rho^{2}C^{2}}\frac{d}{dt}\left(P_{l}-P_{\infty}\right).\nonumber \\
\end{eqnarray}

\noindent The viscous term in Eq'n. (\ref{eq13}) has been
resulted from the simultaneous effects of the liquid viscosity and
compressibility. This term has been neglected in the derived
equations of the previous authors.

To convert Eq'n. (\ref{eq13}) to a differential equation, the
integral term must be approximated. Because of the
irrotationality assumption of the bubble motion the velocity
potential, $\varphi(r,t)$, can be introduced into this equation
as:
\begin{eqnarray}
\label{eq14} \varphi_{t}(R,t) = \frac{\dot
{R}^{2}}{2}+\frac{P_{l}-P_{\infty}}{\rho}
 + \frac{\lambda+2\mu}{\rho^{2}C^{2}}\frac{d}{d
t}\left(P_{l}-P_{\infty}\right),\nonumber \\
\end{eqnarray}

\noindent where subscript $t$ denotes the temporal derivation. In
the simplest approximation of $\varphi_{t}(R,t)$, the liquid
motion is assumed to be incompressible. So, the liquid velocity
at any distance $r$ will be:
\begin{equation}
\label{eq15} u(r)=\frac{R^{2}\dot {R}}{r^{2}}.
\end{equation}

\noindent According to this approximation, $\varphi_{t}(R,t)$ can
be written as:
\begin{equation}
\label{eq16} \varphi_{t}(R,t)=\left[\frac{\partial}{\partial
t}\left(\frac{R^{2}\dot
R}{r}\right)\right]_{r=R}=R\ddot{R}+2\dot{R}^{2}.\nonumber \\
\end{equation}

\noindent Inserting Eq'n. (\ref{eq16}) into Eq'n. (\ref{eq14})
and eliminating the liquid viscous term due to incompressibility
approximation yields the well-known \textit{Rayleigh-Plesset}
equation:
\begin{equation}
\label{eq17}R\ddot{R}+\frac{3}{2}\dot{R}^{2} =
\frac{P_{l}-P_{\infty}}{\rho}.
\end{equation}

To introduced the compressibility effects into approximation of
$\varphi_{t}(R,t)$, we assume that the liquid potential fulfills
the following acoustic equation for the spherical waves:
\begin{equation}
\label{eq18} \frac{\partial}{\partial
t}\left(r\varphi\right)+C\frac{\partial}{\partial
r}\left(r\varphi\right)=0.
\end{equation}

\noindent As we have shown in the following, this approximation
is equivalent to the first order compressibility consideration of
Ref. \cite{Prosperetti:1987}. Double radial differentiation of
Eq'n. (\ref{eq18}) results:
\begin{equation}
\label{eq19} \frac{\partial}{\partial
t}\left(r\triangle\right)+C\frac{\partial}{\partial
r}\left(r\triangle\right)=0.
\end{equation}

\noindent Through the definition of total derivative of the
velocity along with Eq'n. (\ref{eq3}) it can be obtained that:
\begin{eqnarray}
\label{eq19.5}
\varphi_{t}(R,t)\!=\!\!\int_{R}^{\infty}\!\!\frac{\partial
u}{\partial
t}~dr=\int_{R}^{\infty}\!\!\!\left(\frac{du}{dt}-u\triangle+\!\frac{2u^{2}}{r}\right)dr.
\end{eqnarray}
By a partial integration:
\begin{equation}
\label{eq20} \varphi_{t}(R,t)=
R\ddot{R}+2\dot{R}^{2}-R\dot{R}\triangle_{l}
+\!\!\int_{R}^{\infty} r \frac{\partial \triangle}{\partial t}dr.
\end{equation}
Applying Eq'n. (\ref{eq19}) in Eq'n. (\ref{eq20}) yields:
\begin{equation}
\label{eq20.5} \varphi_{t}(R,t)=
R\ddot{R}+2\dot{R}^{2}-R\dot{R}\triangle_{l}+RC\triangle_{l}.
\end{equation}

\noindent Substituting Eq'n. (\ref{eq20}) into Eq'n. (\ref{eq14})
results:
\begin{eqnarray}
\label{eq21}
 R\ddot{R} + \frac{3}{2}\dot{R}^{2} &=& \frac{R}{\rho C}\left(1-\frac{\dot{R}}{C}\right)\frac{dP_{l}}{dt}
 +\frac{P_{l}-P_{\infty}}{\rho} \nonumber\\ &+& \frac{\lambda+2\mu}{\rho^2
C^{2}}\frac{d}{dt}\left(P_{l}-P_{\infty}\right).
\end{eqnarray}

\noindent This equation is the modified form of \textit{Flynn}
equation \cite{Flynn:1975} along with viscosity consideration.

The compressibility effects can be introduced in a different
manner. Differentiating Eq'n. (\ref{eq18}) with respect to $t$
and substituting $\varphi_{t}(R,t)$ in it from Eq'n. (\ref{eq14})
yields:
\begin{eqnarray}
\label{eq22} \frac{R\dot{R}}{C}\left(\frac{\partial u}{\partial t
}\right)_{\!R}+\frac{R}{\rho C}\left(\frac{\partial p}{\partial
t}\right)_{\!R}-\frac{R}{\rho C}\frac{dP_{\infty}}{dt}
+\nonumber \\
R\dot{R}\left(\frac{\partial u}{\partial r}\right)_{\!R}
+\frac{R}{\rho}\left(\frac{\partial p}{\partial
r}\right)_{\!\!R}\!\!+\!\frac{1}{2}\dot{R}^{2}=\nonumber \\
\frac{P_{\infty}-P_{l}}{\rho}-\frac{\lambda+2\mu}{\rho^{2}C^{2}}\frac{d}{dt}(P_{l}-P_{\infty}),
\end{eqnarray}

\noindent in which, the spatial and the temporal partial
derivatives of the viscous term  have been neglected due to
smallness as well as the spatial derivative of the liquid
pressure at infinity. Inserting the relation $d\rho=dp/c^{2}$
into Eq'n. (\ref{eq3}) and using this equation and the momentum
equation along with the total derivatives of the pressure and the
velocity at the bubble interface, the four partial derivatives in
Eq'n. (\ref{eq22}) can be obtained as:
\begin{equation}
\label{eq23} \left(\frac{\partial u}{\partial
t}\right)_{R}=\ddot{R}+\frac{\dot{R}}{\rho
C^{2}}\frac{dP_{l}}{dt}+\frac{2\dot{R}^{2}}{R},
\end{equation}

\begin{equation}
\label{eq24}\left(\frac{\partial p}{\partial t}\right)_{R}=\rho
\dot{R}\ddot{R}+\frac{dP_{l}}{dt},
\end{equation}

\begin{equation}
\label{eq25}\left(\frac{\partial u}{\partial
r}\right)_{R}=-\frac{1}{\rho
C^{2}}\frac{dP_{l}}{dt}-\frac{2\dot{R}}{R},
\end{equation}

\begin{equation}
\label{eq26} \left(\frac{\partial p}{\partial
r}\right)_{R}=-\rho\ddot{R}_{}.
\end{equation}

\noindent Note, in the derivation of the Eq'ns.
(\ref{eq23}-\ref{eq26}), the spatial and the temporal partial
derivatives of the two viscous terms in the Navier-Stokes
equation, Eq'n. (\ref{eq4}), have been neglected. Inserting Eq'ns.
(\ref{eq23}-\ref{eq26}) into Eq'n. (\ref{eq22}) and retaining only
the terms up to the order of $\dot{R}/{C}$ results the modified
form of \textit{Herring-Trilling} equation
\cite{Trilling:1952,Herring:1941} along with viscosity effects:
\begin{eqnarray}
\label{eq27} \left(\!1-\frac{2\dot{R}}{C}\!\right)\!R\ddot{R}\!
\!&\!\!+\!\!&\!\!\left(\!1-\frac{4\dot{R}}{3C}\right)\!\frac{3}{2}\dot{R}^{2}=
\frac{R}{\rho C}\!\frac{d}{dt}\left(P_{l}-P_{\infty}\right)
\nonumber \\
\!&\!+\!&\! \frac{P_{l}-P_{\infty}}{\rho}+\frac{\lambda+2\mu}{
\rho^{2} C^{2}}\frac{d}{dt}\left(P_{l}-P_{\infty}\right).
\nonumber \\
\end{eqnarray}

It is straight forward to obtain the modified form of Eq'n.
(\ref{eq1}) by the addition of Eq'n. (\ref{eq27}) and the product
of $(1-\eta)\dot{R}/C$ with Eq'n. (\ref{eq17}) as:
\begin{eqnarray}
\label{eq28}
\left(1-(\eta+1)\frac{\dot{R}}{C}\right)R\ddot{R}&\!+\!
&\left(1-(3\eta+1)\frac{\dot{R}}{3C}\right)\frac{3}{2}\dot{R}^{2}
=\nonumber\\ \frac{R}{\rho C}\frac{d}{dt}(P_{l}-P_{\infty})
&\!\!+\!\!&\left(1+(1-\eta)\frac{\dot{R}}{C}\right)\frac{P_{l}-P_{\infty}}{\rho}\nonumber
\\&\!+\!&\frac{\lambda+2\mu}
{\rho^{2} C^{2}}\frac{d}{dt}(P_{l}-P_{\infty}).
\end{eqnarray}

The new Eq'n. (\ref{eq28}) provides a suitable description of the
simultaneous effects of liquid compressibility and liquid
viscosity in the bubble motion. The added viscous new term of
this equation arises from the liquid compressibility.

To complete the argument, it is necessary to modify the boundary
condition Eq'n. (\ref{eq2}) with the compressibility effects. The
radial component of the stress tensor is:
\begin{equation}
\label{eq29} T _{rr}=-p+\lambda\nabla\cdot
\vec{u}+2\mu\left(\frac{\partial u }{\partial r}\right).
\end{equation}

\noindent Inserting the velocity divergence from Eq'n.
(\ref{eq3}), into this equation yields:
\begin{eqnarray}
\label{eq30}T_{rr} &=& -p+(\lambda+2\mu)\left(\frac{\partial u
}{\partial r}+\frac{2u}{r}\right)-4\frac{\mu u}{r}\nonumber \\
&=& -p+(\lambda+2\mu)\triangle-4\frac{\mu u}{r}.
\end{eqnarray}
\noindent The boundary continuity requirement at the bubble
interface is:
\begin{eqnarray}
\label{eq31} T_{rr}(liquid)\mid_{R}=
T_{rr}\left(gas\right)\mid_{R}+2\frac{\sigma}{R}.
\end{eqnarray}

\noindent Substituting Eq'ns. (\ref{eq30}) and (\ref{eq12}) into
Eq'n. (\ref{eq31}) leads to:
\begin{eqnarray}
\label{eq32} P_{l}+4\frac{\mu\dot{R}}{R} &+&
\left(\frac{\lambda+2\mu}{\rho C^{2}}\frac{dP_{l}}{dt}\right)=
P_{g}+ 4\frac{\mu_{g}\dot{R}}{R}
 \nonumber \\ &+&
\left(\frac{\lambda_{g}+2 \mu_{g}}{\rho_{g}}\frac{d
\rho_{g}}{dt}\right)-2\frac{\sigma}{R},
\end{eqnarray}

\noindent where, $\rho_{g}$, $\mu_{g}$, and $\lambda_{g}$ are the
gas density, the first and the second coefficients of viscosity
of the gas, respectively. The new Eq'n (\ref{eq32}) provides the
most complete boundary condition equation at the bubble
interface, which contains all effects of compressibility and
viscosity of both the liquid and the gas. Here, we concentrate on
the effects of liquid viscous new term of Eq'n (\ref{eq32}) and
eliminate the gas viscosity terms as previous works
\cite{Rayleigh:1917,Plesset:1949,Noltingk:1951,
Herring:1941,Trilling:1952,Gilmore:1952,Keller:1956,
Hickling:1963,Fujikawa:1980,Yasui:1997,Hickling:1964,Jahsman:1968,
Flynn:1975,Lastman:1979,fstedt:1993,Nigmatulin:2000,Brujan:2001,
Prosperetti:1987,Keller:1980}. Therefore, Eq'n. (\ref{eq32})
becomes:
\begin{eqnarray}
\label{eq33} P_{l}+\left(\frac{\lambda+2\mu}{ \rho
C^{2}}\frac{dP_{l}}{dt}\right)=P_{g} -4\frac{
\mu\dot{R}}{R}-2\frac{\sigma}{R}.
\end{eqnarray}

\noindent Comparison of Eq'ns. (\ref{eq2}) and (\ref{eq33})
indicates the existence of a new viscous term in Eq'n.
(\ref{eq33}) due to the liquid compressibility. The set of Eq'ns.
(\ref{eq28}) and (\ref{eq33}) present the most complete form of
the bubble dynamics equations containing effects of the liquid
compressibility and viscosity not only at the bubble interface,
but also in the bulk of liquid. In fact, these equations account
for the viscosity of a compressible liquid. While, all previous
equations accounted for the viscosity of an incompressible liquid
and compressibility, separately.

To generalize the argument, Eq'ns. (\ref{eq28}) and (\ref{eq33})
are expressed in dimensionless forms. The dimensionless variables
of this problem are defined as:
\begin{eqnarray}
\label{eq34} R^{\ast}=\frac{R}{R_{0}}, &\;&
\dot{R^{\ast}}=\frac{\dot{R}}{C},\; \; \; \; t^{\ast}=\frac{t
C}{R_{0}}, \; \; \; \;P_{l}^{\ast}=\frac{P_{l}}{\rho
C^{2}}, \; \; \; \; \nonumber \\
&& P_{g}^{\ast}=\frac{P_{g}}{\rho C^{2}}, \; \; \; \;
P_{\infty}^{\ast}=\frac{P_{\infty}}{\rho C^{2}}, \; \; \; \;
\end{eqnarray}

\noindent where, $R_{0}$ is ambient radius of the bubble.
Substituting the dimensionless variables into Eq'ns. (\ref{eq28})
and (\ref{eq33}), the dimensionless equations are obtained as:
\begin{eqnarray}
\label{eq35}\left(1-(\eta+
1)\dot{R}^{\ast}\right)\!\!\!&\!\!\!\!\!\!\!\!&\!\!\!\!\!\!
R^{\ast} \ddot{R^{\ast}}+\frac{3}{2}\left(1-\frac{1}{3}(3\eta
+1)\dot{R}^{\ast}\right)\dot{R}^{\ast 2} \nonumber
\\ &=&\!\left(\!{1+(1-\eta)\dot{R}^{\ast}
}\!\right)\!\!\left(P_{l}^{\ast}-P_{a}^{\ast}-P_{0}^{\ast}\right)
\nonumber\\&+&
R^{\ast}\frac{d}{dt^{\ast}}\!\left(P_{l}^{\ast}-P_a^{\ast}\right),
\end{eqnarray}

\begin{equation}
\label{eq36}
P_{l}^{\ast}=P_{g}^{\ast}-4\frac{\mu^{\ast}\dot{R}^{\ast}}{R^{\ast}}-2\frac{\sigma^{\ast}}{R^{\ast}}
-(\lambda^{\ast}+2\mu^{\ast})\frac{dP_{l}^{\ast}}{dt^{\ast}}.
\end{equation}
\noindent The quantities $\sigma^{\ast}$, $\mu^{\ast}$, and
$\lambda^{\ast}$ are dimensionless surface tension and
dimensionless liquid viscosity coefficients, which are defined as:
$\sigma^{\ast}=\sigma/\rho R_{0}C^{2}$, $\mu^{\ast}=\mu/\rho
R_{0}C$, and $\lambda^{\ast}=\lambda/\rho R_{0}C$. These
dimensionless numbers, which are basically inverse of Weber
Number and inverse of Reynolds Number, characterize significance
of the surface tension and the liquid viscosity in the bubble
dynamics.

\section{Bubble  Interior  Evolution}

To quantify effects of the new viscous terms on the bubble
dynamics, evolution of the gas pressure at the bubble interface,
$P_{g}$, must be specified. It can be determined from
simultaneous solution of the conservation equations for the
bubble interior and the bubble radius equations \cite{C.C.WU,
Moss:1994, Kondict:1995, Voung:1996, Yuan:1998, Xu:2003}. Also,
heat conduction and mass exchange between the bubble and the
surrounding liquid affect the bubble evolution. These
complexities were considered in a complete gas dynamics
simulation by Storey and Szeri \cite{Storey:2000}.

On the other hand, strong spatial inhomogeneities inside the
bubble are not remarkably revealed, unless at the end of an
intense collapse \cite{Voung:1996, Yuan:1998}. Therefore, the
uniformity assumption for the bubble interior seems to be useful
and provides many features of the bubble motion
\cite{Barber:1997, Brenner:2002}. Using this assumption Toegel
\textit{et al.} presented an ODE model \cite{Toegel:2000}, in
which effects of heat and mass transfer at the bubble interface
have been considered. This model accurately describes various
experimental phase diagrams \cite{Toegel:2003} and provides a
good agreement with the complete direct numerical simulation of
Storey and Szeri \cite{Storey:2000}.

In this paper, we used the Toegel \textit{et al.}'s model
\cite{Toegel:2000} for specifying the bubble interior evolution.
We describe an argon bubble in water under the conditions of
Single Bubble Sonoluminescence \cite{Brenner:2002, Barber:1997}.
The gas evolution model can be summarized as follows.

The gas pressure is modeled by the van der Waals equation of
state:
\begin{equation}
\label{eq37} P_{g}=\frac{N_{tot}kT}{V-N_{tot}B}
\end{equation}
where, $N_{tot}=N_{Ar}+N_{H_{2}O}$ is total number of particles
inside the bubble. The covolume constant
$B=5.1\times10^{-29}~m^{3}$ is assumed to be equal for both water
vapor and argon \cite{Toegel:2003}. The value of $N_{tot}$
changes with time because of evaporation and condensation of the
water vapor molecules at the bubble interface. The rate of change
can be modeled as \cite{Toegel:2000}:
\begin{eqnarray}
\label{eq38} \dot{N}_{tot}\!=\!4\pi R^{2} D
\frac{n_{0_{H_2O}\!}\!-\!n_{_{H_2O}}}{l_d},~~~~
l_d=min\!\left(\!\sqrt{\frac{R
D}{|\dot{R}|}},\!\frac{R}{\pi}\right)\!, \nonumber
\\
\end{eqnarray}

\noindent where, $n_{_{H_2O}}$ and $n_{0_{H_2O}}$ are the
instantaneous and equilibrium concentration of the water vapor
molecules, respectively. The diffusion coefficient, D, is given
by \cite{Lu:2003}: $D=D_0(n_{0}/n_{tot})$, where
$D_0=23.55\times10^{-6}~m^2/s$ and $n_0= 2.5
\times10^{25}~m^{-3}$. The quantity $l_d$ is thickness of
diffusive boundary layer (see Ref. \cite{Toegel:2000} for more
details). The equilibrium concentration of the water vapor
molecules is given by the number density of the saturated vapor
pressure at the ambient liquid temperature $T_0$;
$n_{0_{H_2O}}=P_{\nu}(T_{0})/kT_{0}$. In our calculations:
$n_{0_{H_2O}}=5.9\times10^{23}~m^{-3}$.

Similar to Eq'n. (\ref{eq38}), the heat exchange at the bubble
interface can be approximate by \cite{Toegel:2000}:
\begin{equation}
\label{eq39}\dot{Q}=4 \pi R^{2}\kappa \frac{T_{0}-T} {l_{th}},
~~~~ l_{th}=min\left(\sqrt{\frac{R
\chi}{|\dot{R}|}},\frac{R}{\pi}\right),
\end{equation}
where, $T$ and $T_0$ are the gas temperature and the ambient
liquid temperature, respectively. Also, $\kappa=17.9\times10^{-3}~
W/mK$ \cite{Lu:2003} is thermal conductivity coefficient of the
gas content, $l_{th}$ is thickness of thermal boundary layer and
$\chi$ is thermal diffusivity coefficient, which is given by:
$\chi=\kappa/c_p$, with
$c_p=(\frac{5}{2}n_{Ar}+\frac{5}{2}n_{H_2O})k$ the constant
pressure heat capacity per unit volume of the gas.

Applying the basic energy equation for the bubble content
evolution results the rate of change of the gas temperature
\cite{Toegel:2000,Toegel:2003}:
\begin{equation}
\label{eq40}
\dot{T}=\frac{\dot{Q}}{C_{\nu}}-\frac{P_g\dot{V}}{C_{\nu}}+\!\left[4T_0-\!
3T-\!T\!\sum\!\left(\frac{\theta_i/T}{e^{\theta_i/T}-1}\right)\right]\!
\frac{k\dot{N}_{tot}}{C_{\nu}}
\end{equation}
\begin{eqnarray}
\label{eq41}
C_{\nu}=\frac{3}{2}N_{Ar}k+\!\!\left[3+\!\sum\!\left(\frac{(\theta_i/T)^2e^{\theta_i/T}}
{(e^{\theta_i/T}-1)^2}\right)\!\right]k N_{H_2O},
\end{eqnarray}

\noindent where, three different values for $\theta_i$ correspond
to characteristic vibration temperatures of H$_2$O:
$\theta_1=2295 K$, $\theta_2=5255 K$, and $\theta_3=5400 K$
\cite{Toegel:2003}. In this paper for simplicity we neglected the
effects of chemical reactions \cite{Toegel:2003}, which are
important only at an extremely short time interval when the
bubble temperature is high enough (more than 5000K).
\begin{figure}[t]
\vskip 5mm
\includegraphics[width=8cm,height=5cm]{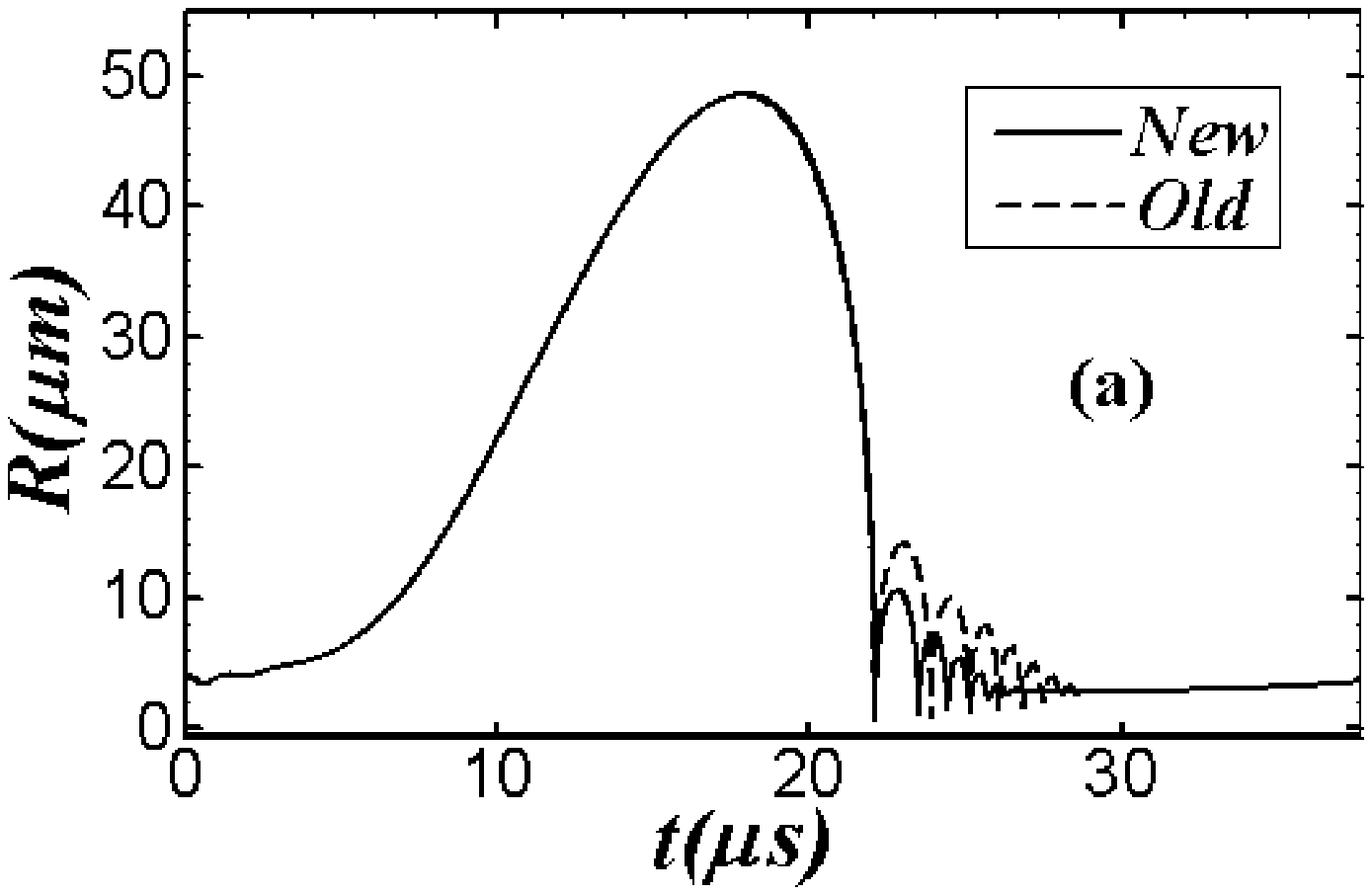}
\vskip 10mm
\includegraphics[width=8cm,height=5cm]{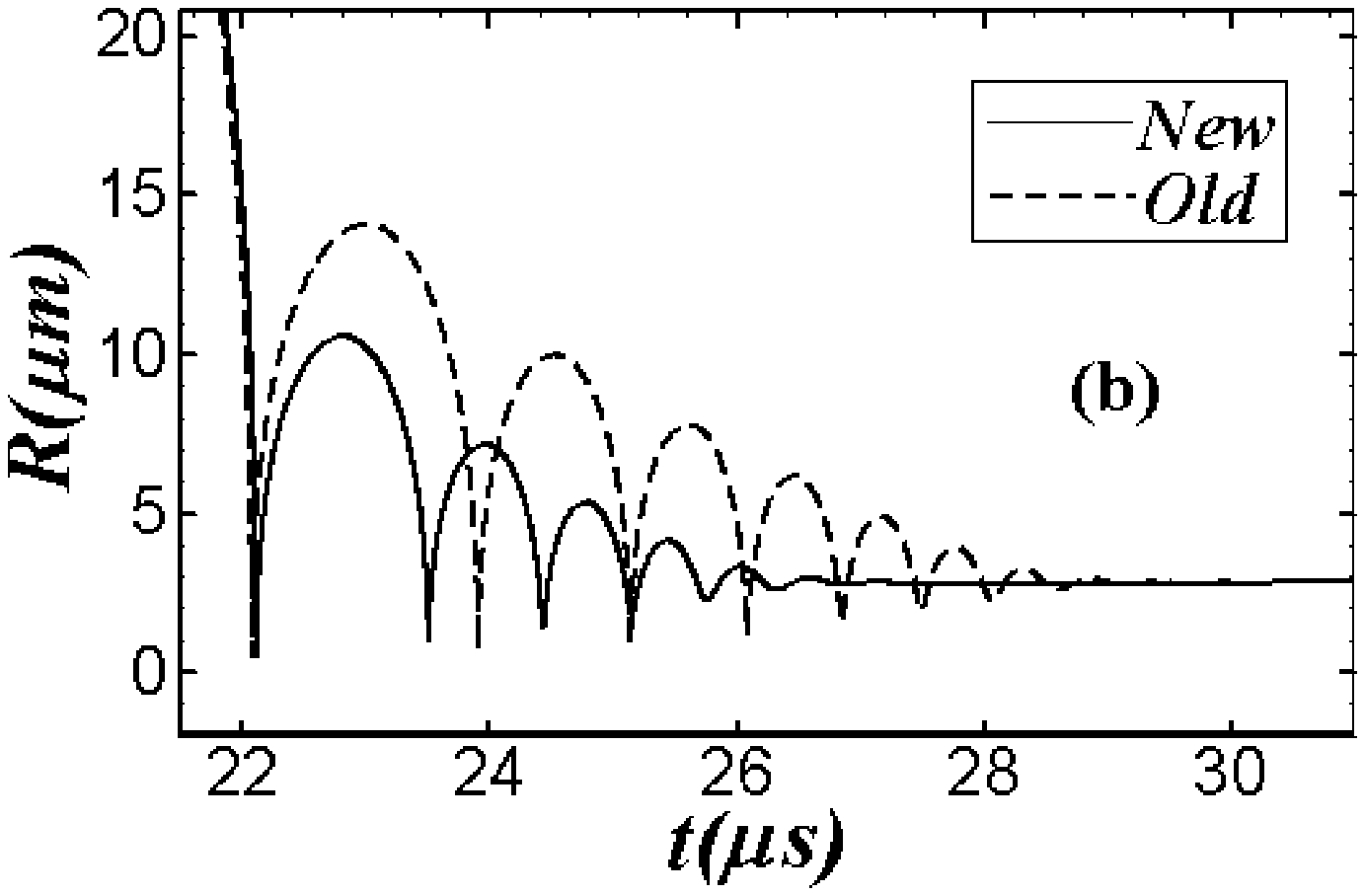}
\caption{ Time variations of the bubble radius, according to the
new (solid) and the old (dashed) sets of equations. Graphs (a)
and (b) shows the bubble evolution in a complete period and
during the bubble rebounds, respectively. The equilibrium radius
is $R_{0}=4.0 ~\mu m$ and the deriving pressure amplitude is
$P_{a}=1.4 ~atm$}. \label{fig1:dls}
\end{figure}

Equations (\ref{eq37}-\ref{eq41}) along with the bubble dynamics
equations are the set of equations, which totally describe the
evolution of the bubble characteristics. Under these
circumstances, time variations of the bubble properties have been
numerically calculated for both the new (Eq'ns.
\ref{eq35},\ref{eq36}) and the old (Eq'ns. \ref{eq1},\ref{eq2})
bubble dynamics equations (for $\eta=0$).

\section{Numerical Analysis}

The calculations were carried out for a periodic driving pressure;
$P_{a}(t)=P_{a}\sin \left(\omega t\right)$, with
$\omega=2\pi\times26.5~kHz$. The constants and parameters of the
bubble dynamics equations were set for the water at room
temperature, $T_{0}=293.15~K$, and atmospheric ambient pressure,
$P_{0}=1.0 ~atm$; $\rho=998.0 ~kg/m^{3}$, $C=1483.0 ~m/s$,
$\mu=1.01\times10^{-3} ~kg/ms$, $\sigma=0.0707 ~kgs^{-2}$
\cite{CRC:1995}. The second coefficient of viscosity of water at
room temperature was set to be $\lambda=3.43\times10^{-3} ~kg/ms$
\cite{Karim:1952}.

Figures (1-3) illustrate the results of our calculations for
$P_{a} = 1.4 ~atm$ and $R_{0}=4.0 ~\mu m$. Similar values for
these parameters have been reported in recent experimental works
of Ketterling and Apfel \cite{Ketterling:2000}, Simon \textit{et
al.} \cite{Simon:2001}, and Vazquez \textit{et al.}
\cite{Vazquez:2002}. Figure (1) shows the variations of the
bubble radius for the new and the old bubble dynamics equations.
It is observed that the addition of the new viscous terms
considerably changes the bubble evolution after the collapse. The
bubble motion is remarkably compressible during the collapse.
Therefore, the new viscous terms, which have been arisen from the
liquid compressibility, are important in this time interval.
These terms exhibit a damping role and their consideration
reduces the amplitude of the bubble rebounds. Also, the period of
the rebounds decreases with the addition of the new terms.
Details of our calculations at the end of the collapse show that
the minimum radius for the new case is about 4\% greater than
that of the old one.
\begin{figure}[t]
\vskip 4mm
\includegraphics[width=8cm,height=5cm]{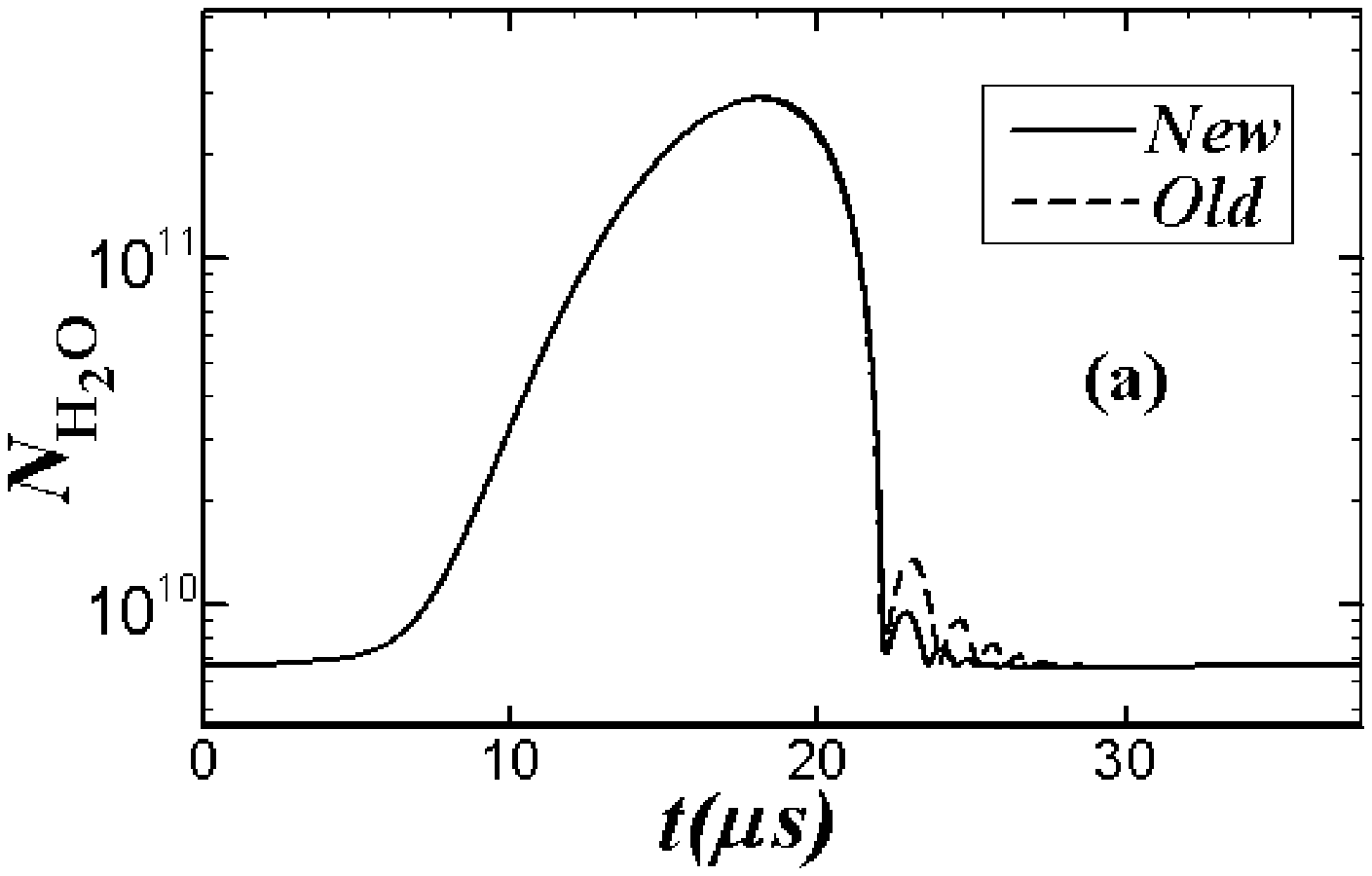}
\vskip 10mm
\includegraphics[width=8cm,height=5cm]{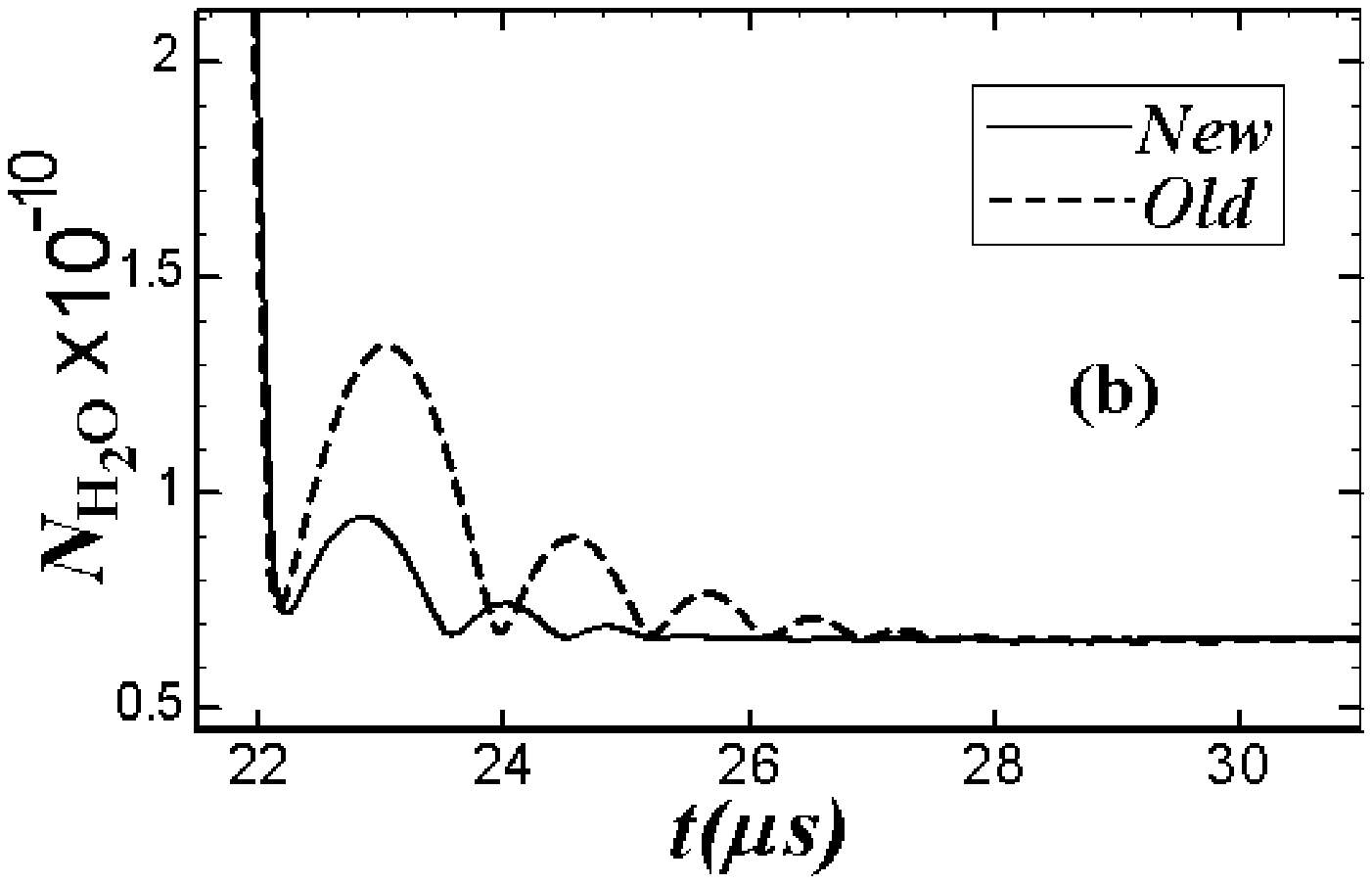}
\caption{Time variations of the number of H$_2$O particles inside
the bubble according to the new (solid) and the old (dashed) sets
of equations. Graphs (a) and (b) show the evolution in a complete
period and during the bubble rebounds, respectively. The
parameters and constants are the same as Fig. (1).}
\label{fig2:dls}
\end{figure}
\begin{figure}[t]
\vskip 4mm
\includegraphics[width=8cm,height=5cm]{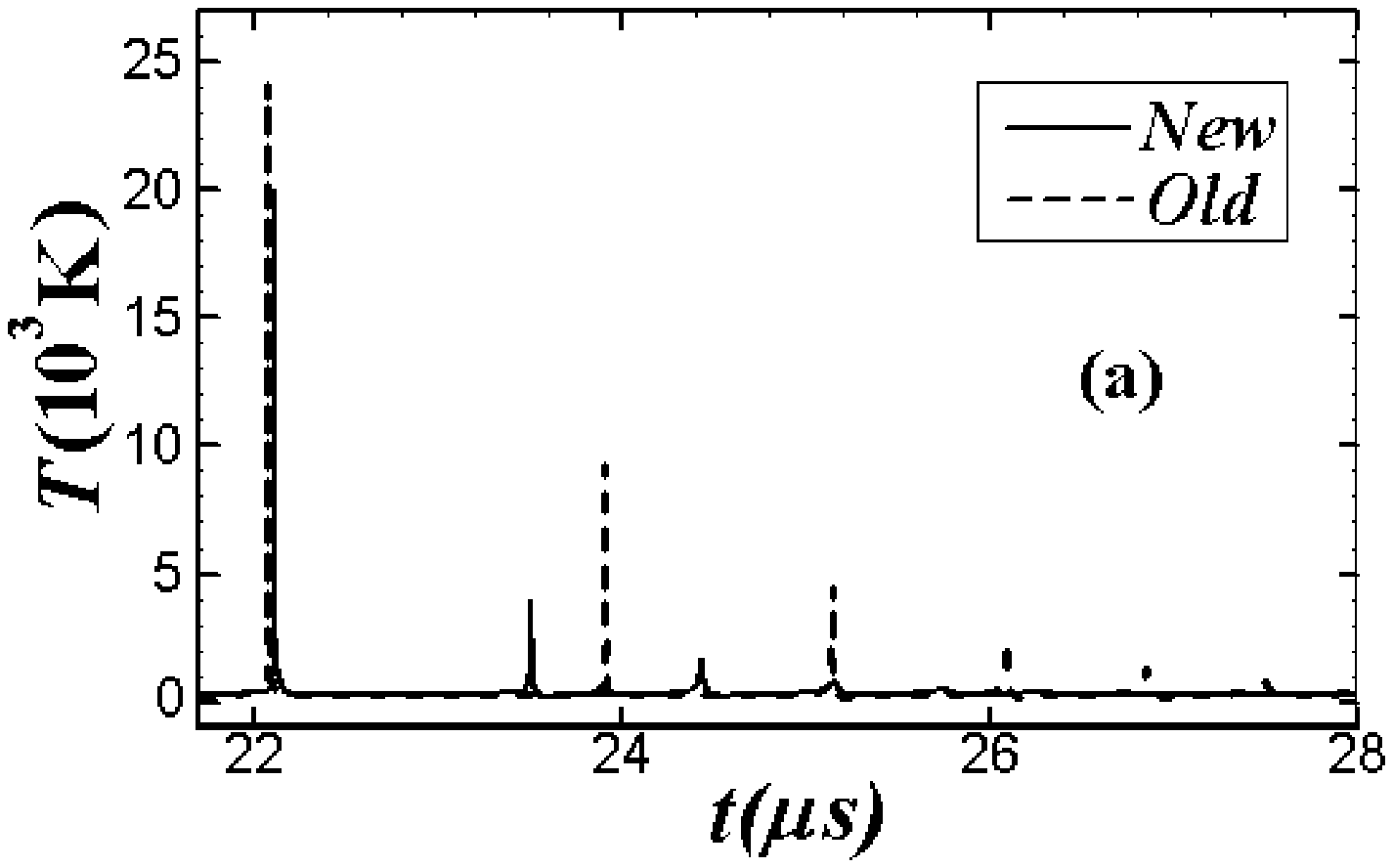}
\vskip 10mm
\includegraphics[width=8cm,height=5cm]{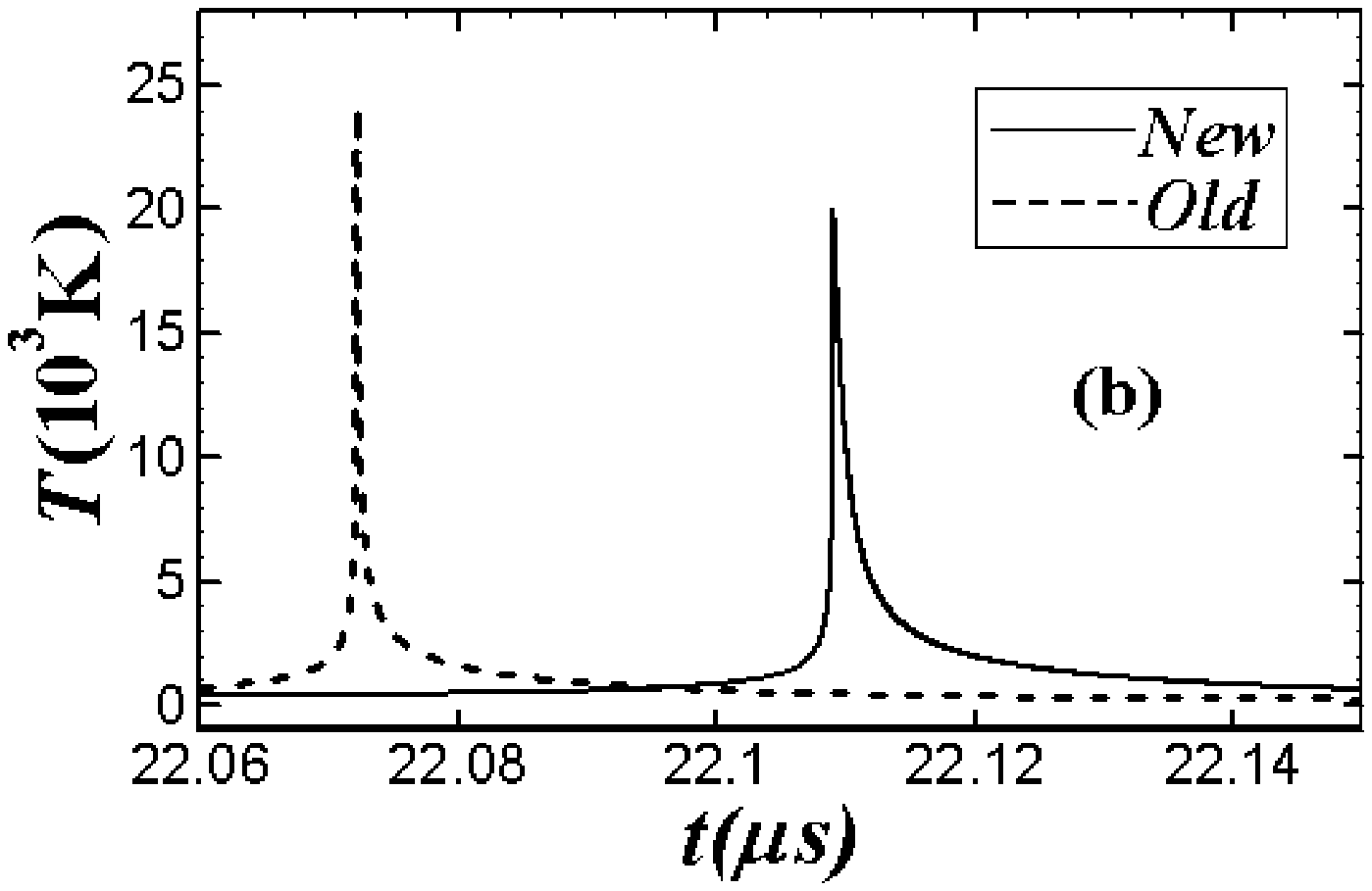}
\caption{The gas temperature evolution during the bubble rebounds
(a) and at the end of the collapse (b) for the the new (solid)
and the old (dashed) cases for the same parameters and constants
as Fig. (1).} \label{fig3:dls}
\end{figure}

\begin{figure}[t]
\vskip 4mm
\includegraphics[width=8cm,height=4.9cm]{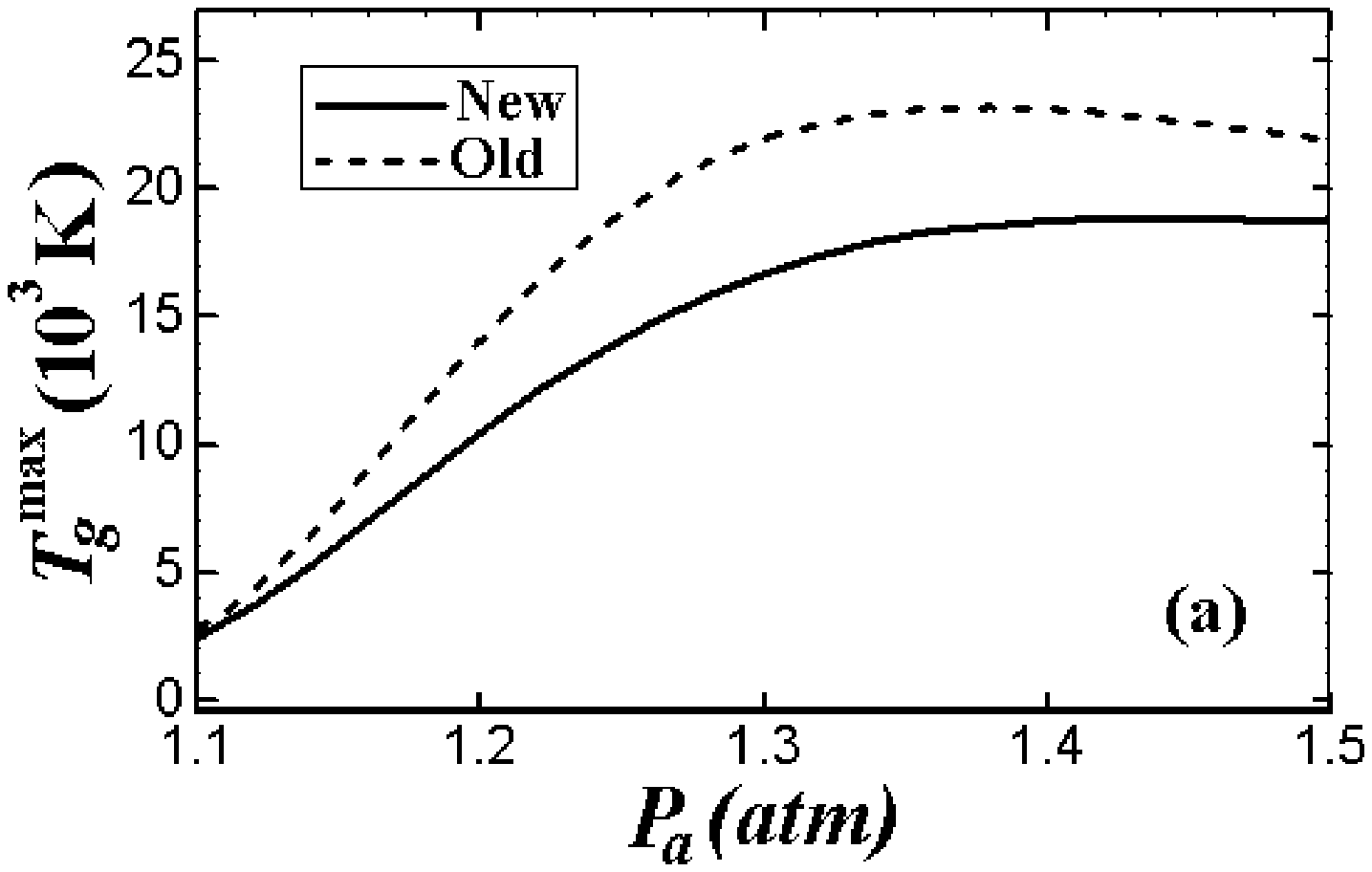}
\vskip 10mm
\includegraphics[width=8cm,height=4.9cm]{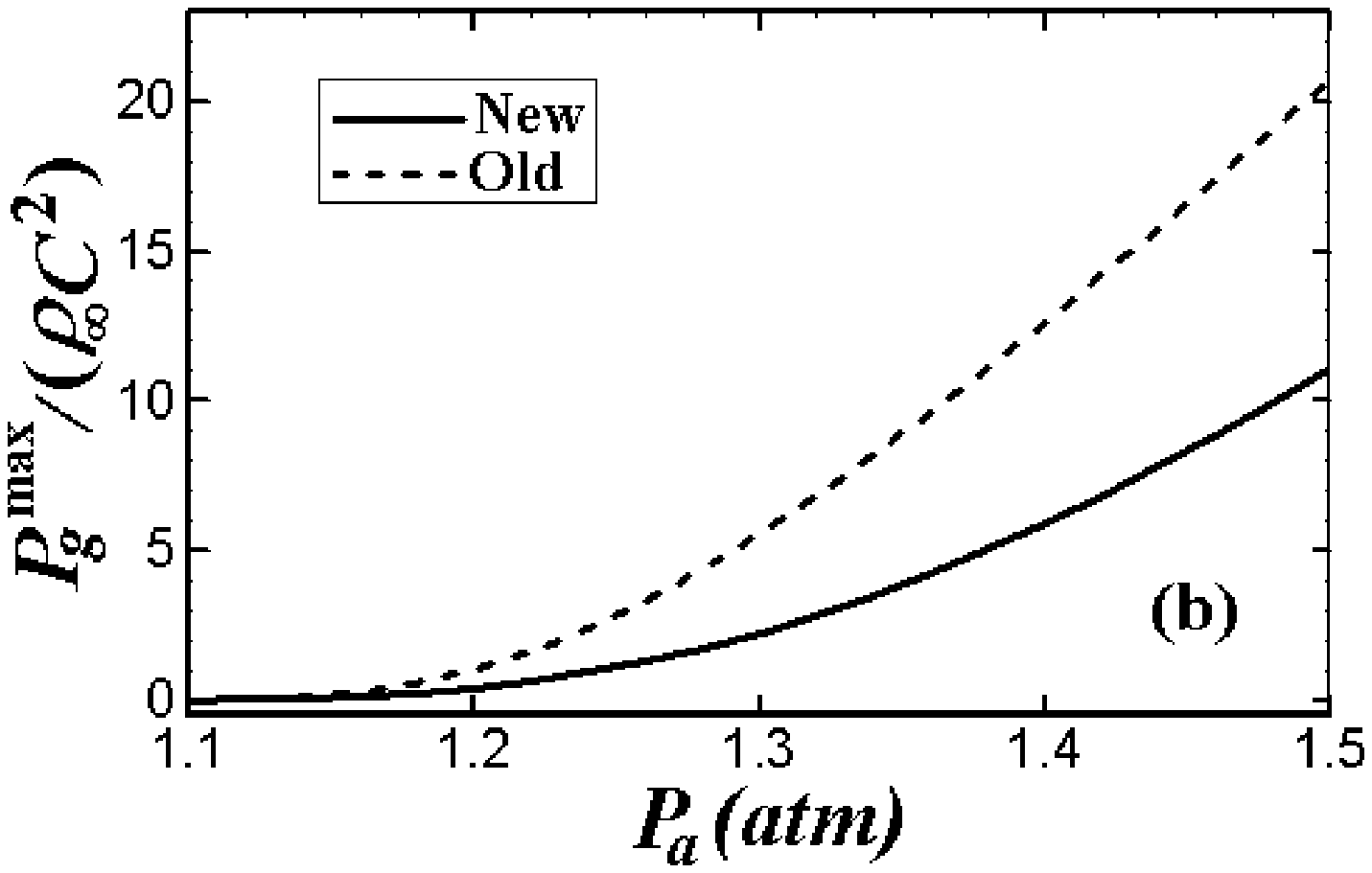}
\vskip 10mm
\includegraphics[width=8cm,height=4.9cm]{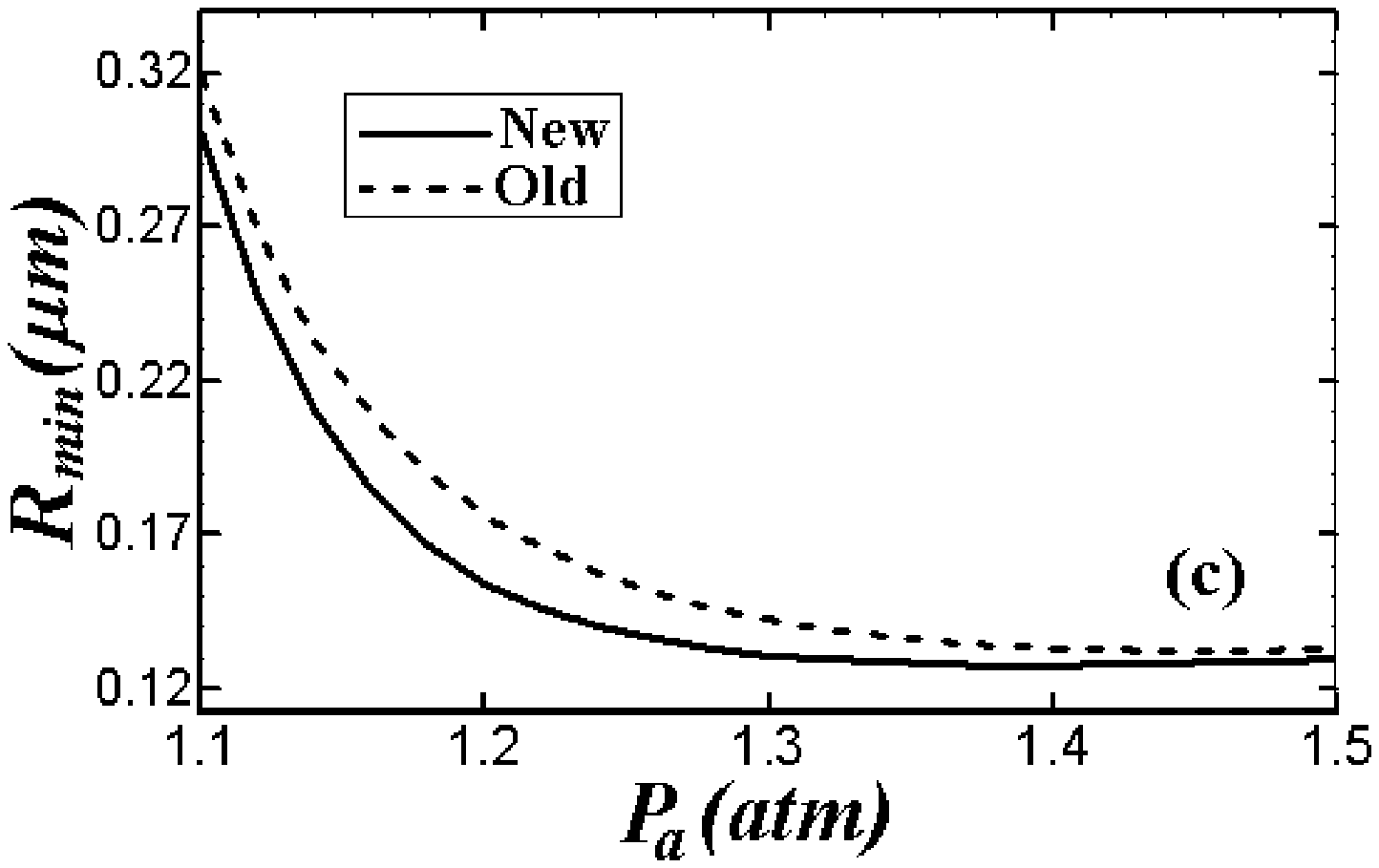}
\caption{ The bubble characteristics at the time of collapse as a
function of the driving pressure amplitude for the new (solid)
and the old (dashed) bubble dynamics equations; peak temperature
(a), peak pressure (b), and minimum radius (c). The equilibrium
radius was fixed ($R_0=4.5 ~\mu m$). Other constants are the same
as Figs. (1-3).} \label{fig4:dls}
\end{figure}

\begin{figure}[t]
\vskip 4mm
\includegraphics[width=8cm,height=4.9cm]{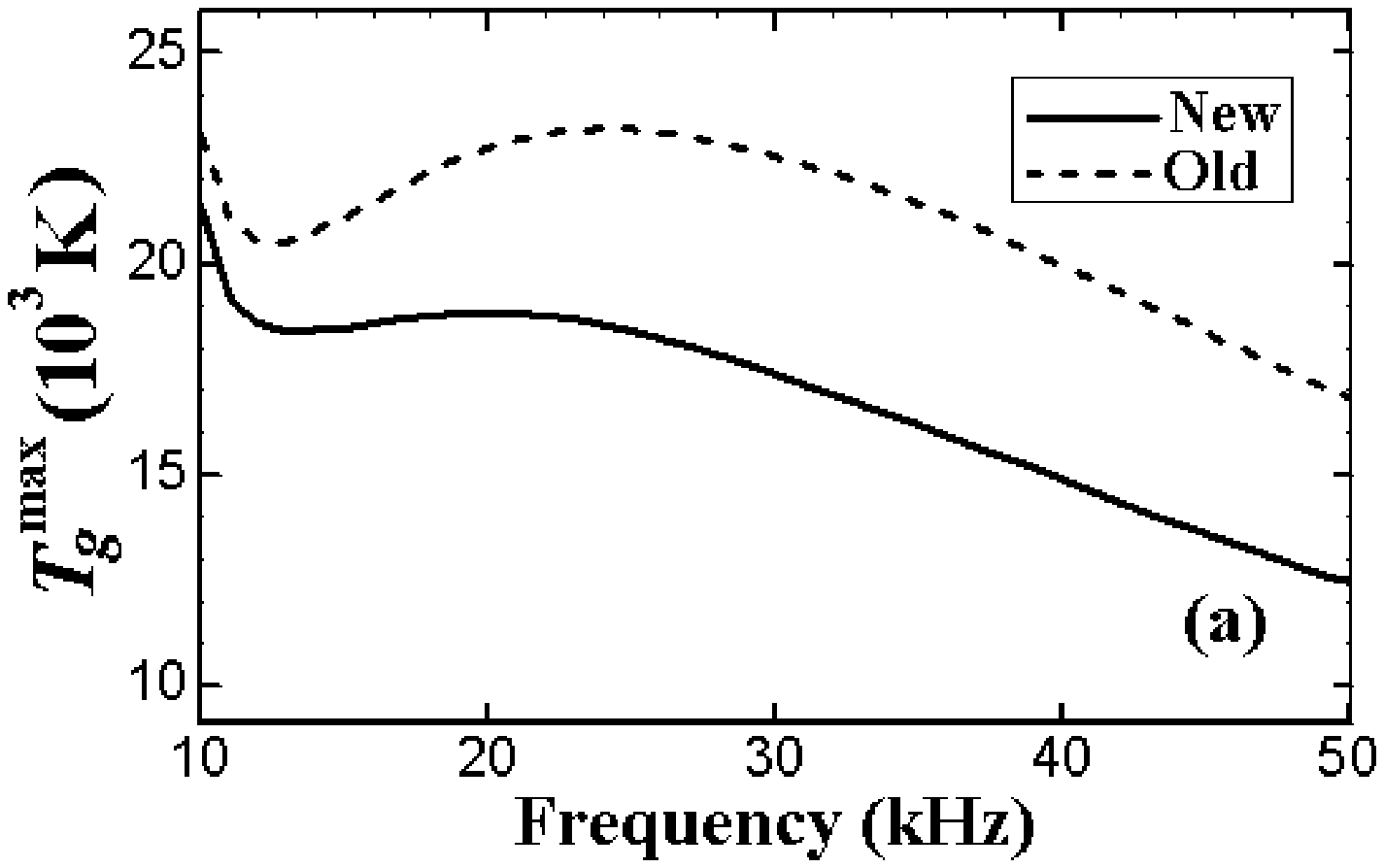}
\vskip 10mm
\includegraphics[width=8cm,height=4.9cm]{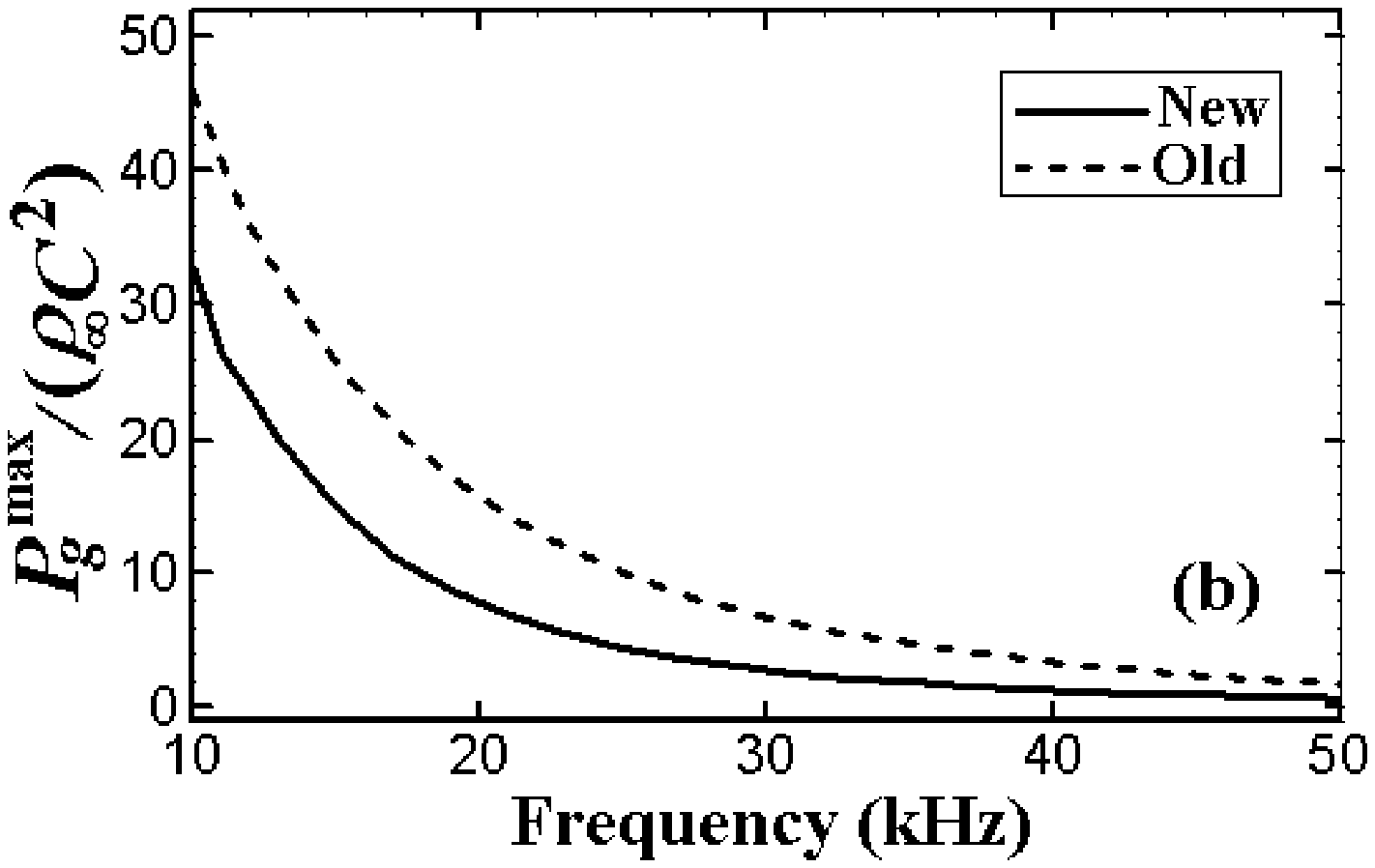}
\vskip 10mm
\includegraphics[width=8cm,height=4.9cm]{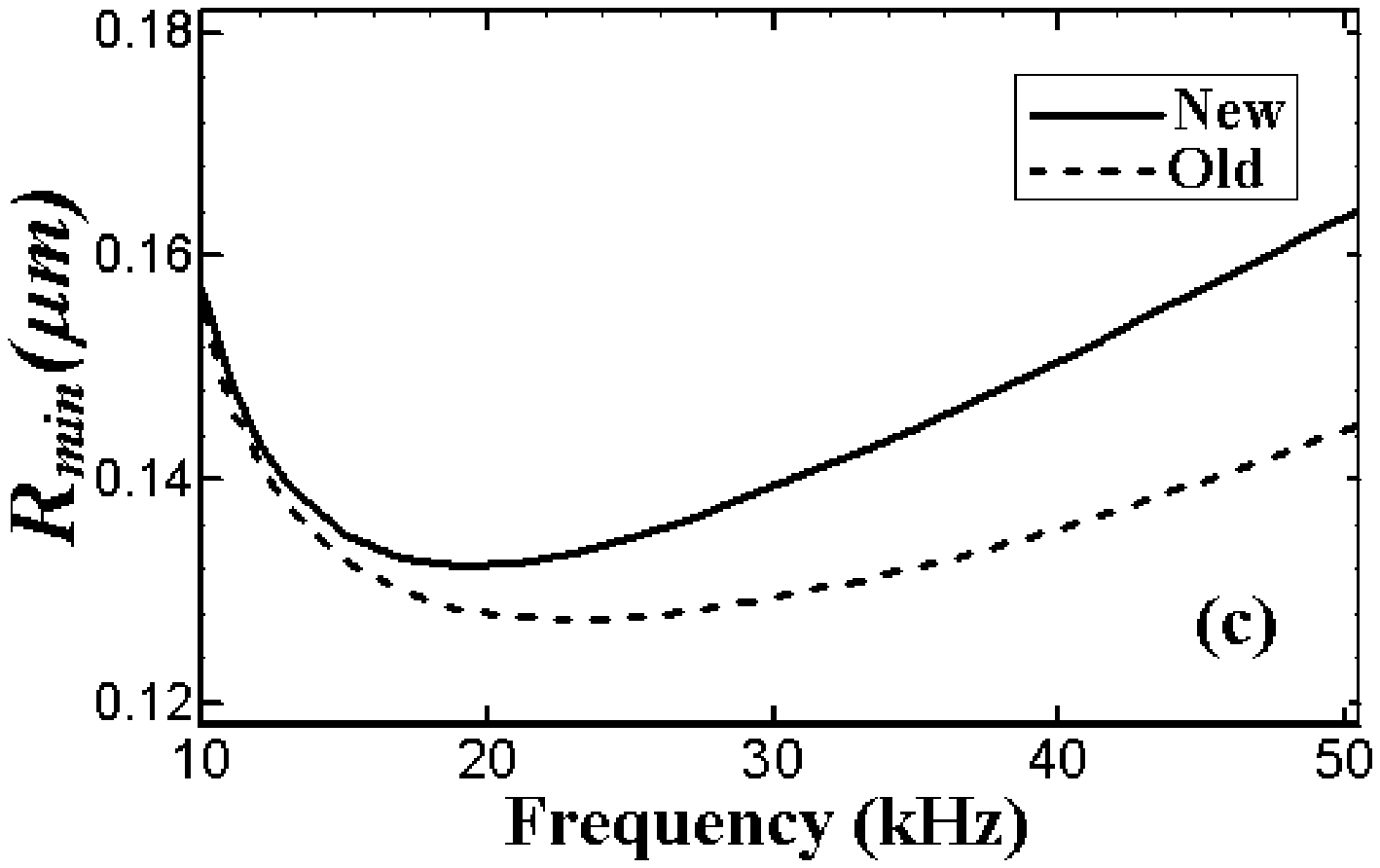}
\caption{ The dependence of the bubble characteristics at the end
of the collapse to the frequency of driving pressure for the new
(solid) and old (dashed) bubble dynamics equations; peak
temperature (a), peak pressure (b), and minimum radius (c). The
equilibrium radius and the pressure amplitude were fixed ($R_0=4.5
~\mu m$ and $P_{a} = 1.35 ~atm$). Other constants are the same as
Figs. (1-3).} \label {fig5:dls}
\end{figure}

The difference between the two cases for the evolution of the
number of H$_2$O particles has been illustrated in Fig. (2). As
this figure indicates, a large amount of water vapor evaporates
into the bubble during the expansion. Indeed, at maximum radius,
most of the bubble content is water vapor. During the collapse,
the water vapor molecules rapidly condense to the liquid. It is
seen that the remarkable difference between the two cases also
appears on the variations of the number of H$_2$O particles after
the collapse. Note that, the difference gradually disappears as
the bubble rebounds weaken.

In Fig. (3), the gas temperature evolution during the bubble
rebounds and at the end of the collapse have been demonstrated.
Damping feature of the new term is clearly observed by
considerable decrease of the peak temperature at the collapse
time as well as remarkable reduce of the secondary peak
temperatures. Also, the pulse width of the main peak temperature
decreases with the addition of the new terms.

Figures (4) and (5) illustrate dependence of the bubble
properties at the end of the collapse to the variation of the
amplitude and the frequency of driving pressure for the two
cases. The different values of $P_{a}$, corresponding to a
specific values of $R_{0}$, can be experimentally obtained by
varying the concentration of the dissolved gas in the liquid
\cite{Barber:1997,Ketterling:2000,Simon:2001,Vazquez:2002}. Also,
the frequency range in the calculations of Fig. (5) is in the
range of the experimental results \cite{Barber:1997}.

Figure (4) shows the variation of the peak temperature, the peak
pressure, and the minimum radius as a function of the driving
pressure amplitude. The ambient radius was fixed ($R_0=4.5 ~\mu
m$). The damping feature of the new viscous terms is clearly
observed in this figure. For the peak temperature in Fig. 4(a),
the maximum relative difference appears around $P_a=1.25~atm$
(about 35\%). The difference between the two cases decreases for
higher driving pressures. However, for the peak pressure in Fig.
4(b), the difference increases with the increase of the
amplitude. The values of $R_{min}$ for the two cases are similar
in high driving pressures because of the effects of the excluded
volume, which prevent the bubble compression \cite{Toegel:2002}.

Figure (5) represents the dependence of the bubble characteristics
at the end of the collapse to the variations of the driving
pressure frequency. The ambient radius and the amplitude were
fixed in these calculations ($R_0=4.5 ~\mu m$ and
$P_a=1.35~atm$). As Fig. (4), the damping feature of the new terms
is also seen here. The difference between the two cases for the
peak temperature and the minimum radius increases when the
frequency is increased. While, for the peak pressure the
difference reduces with the increase of the frequency.

A major deficiency of the old bubble dynamics equations is that
for strongly driven bubbles, such as sonoluminescence bubbles,
large amplitude rebounds are produced after the collapse, so that
they often last until the next acoustic cycle of the periodic
driving pressure. This is in contrast with the experimental
results, which show rapidly damped rebounds
\cite{Barber:1997,Moss:2000}. By introducing a damping term
arisen from the gas compressibility, Moss \textit{et. al}
\cite{Moss:2000} provided a typical solution for this problem.
The effects of the suggested term by Moss \textit{et. al} is very
similar to the damping effects of the new terms in this paper,
(compare Fig. 1(b) with Figs. (3) and (4) of Ref.
\cite{Moss:2000}). It seems the damping feature of the the new
terms is a better way for solving the mentioned problem. The
reason is that Eq'ns. (\ref{eq35}) and (\ref{eq36}) have been
derived directly from the basic equations of fluid mechanics, on
the contrary to Eq'n. (3.2) of Ref. \cite{Moss:2000}, which was
derived by an approximate method.

\section{Conclusions}

The equations of the bubble motion in a compressible viscous
liquid were newly derived from the full Navier-Stokes equations.
These equations contain two similar new terms resulted from the
simultaneous effects of the liquid viscosity and compressibility.
These new terms have a considerable damping role at the collapse,
when the bubble motion is significantly compressible. This new
damping mechanism dramatically changes the bubble properties at
the end of the collapse and during the bubble rebounds.

The results of this work indicate that, the neglect of the new
terms in the previous works is not reasonable for the collapse
time and the new effects should be considered for the prediction
of the quantities related to the collapse, e.g., the value of
light emission by a single sonoluminescing bubble as well as the
bubble stability limits.

\section*{ACKNOWLEDGEMENTS}

This work was supported by Sharif University of Technology and
Bonab Research Center. Partial support of this work by Institute
for Studies in Theoretical Physics and Mathematics is
appreciated. The authors thank Andrea Prosperetti for his helpful
comments.

\end{document}